\def\a{\alpha}\def\b{\beta}\def\d{\delta}
\def\f{\phi}\def\g{\gamma}
\def\k{\kappa}\def\l{\lambda}\def\m{\mu}\def\n{\nu}\def\q{\psi}\def\r{\rho}\def\s{\sigma}
\def\y{\eta}

\def\D{\Delta}\def\F{\Phi}

\def\de{\partial}
\def\id{\equiv}\def\mo{{-1}}\def\ha{{1\over 2}}
\def\qu{{1\over 4}}

\def\({\left(}\def\){\right)}\def\[{\left[}\def\]{\right]}
\def\lra{\leftrightarrow}
\def\mn{{\mu\nu}}\def\ij{{ij}}

\def\tr{\triangleright}

\def\cor{commutation relations }\def\poi{Poincar\'e }\def\coo{coordinates }
\def\ie{i.e.\ }

\def\section#1{\bigskip\noindent{\bf#1}\smallskip}

\def\nota{\footnote{$^\dagger$}}

\def\PL#1{Phys.\ Lett.\ {\bf#1}}\def\CMP#1{Commun.\ Math.\ Phys.\ {\bf#1}}

\def\PR#1{Phys.\ Rev.\ {\bf#1}}\def\CQG#1{Class.\ Quantum Grav.\ {\bf#1}}

\def\JMP#1{J.\ Math.\ Phys.\ {\bf#1}}

\def\JoP#1{J.\ Phys.\ {\bf#1}} \def\IJMP#1{Int.\ J. Mod.\ Phys.\ {\bf #1}}
\def\MPL#1{Mod.\ Phys.\ Lett.\ {\bf #1}}

\def\JHEP#1{JHEP\ {\bf#1}}\def\EPJ#1{Eur.\ Phys.\ J.\ {\bf#1}}

\def\ref#1{\medskip\everypar={\hangindent 2\parindent}#1}
\def\beginref{\begingroup
\bigskip
\centerline{\bf References}
\nobreak\noindent}
\def\endref{\par\endgroup}

\def\hx{\hat x}\def\hm{\hat m}
\def\cP{{\cal P}}\def\cK{{\cal K}}\def\cD{{\cal D}}\def\cF{{\cal F}}
\def\ab{{\a\b}}\def\rs{{\r\s}}\def\ot{\otimes}
\def\bl{{\bf l}}\def\br{{\bf r}}

%\magnification=1200
%\baselineskip18pt
%\input macro.tex
{\nopagenumbers
\line{}
\vskip40pt
\centerline{\bf Associative realizations of the extended Snyder model}
\vskip40pt
\centerline{{\bf S. Meljanac}\nota{e-mail:meljanac@irb.hr}}
\vskip5pt
\centerline {Rudjer Bo\v skovi\'c Institute, Theoretical Physics Division}
\centerline{Bljeni\v cka c. 54, 10002 Zagreb, Croatia}
\vskip10pt
\centerline{and}
\vskip5pt
\centerline{{\bf S. Mignemi}\nota{e-mail:smignemi@unica.it}}
\vskip5pt
\centerline {Dipartimento di Matematica, Universit\`a di Cagliari}
\centerline{via Ospedale 72, 09124 Cagliari, Italy}
\smallskip
\centerline{and INFN, Sezione di Cagliari}

\vskip40pt
{\noindent\centerline{\bf Abstract}
}
\vskip60pt
The star product usually associated to the Snyder model of noncommutative geometry is nonassociative,
and this property prevents the construction of a proper Hopf algebra. It is however possible to introduce a
well-defined Hopf algebra by including the Lorentz generators and their conjugate momenta into the algebra.
In this paper, we study the realizations of this extended Snyder spacetime, and obtain the coproduct
and twist and the associative star product in a Weyl-ordered realization, to first order in the noncommutativity
parameter.
We then extend our results to the most general realizations of the extended Snyder spacetime, always up to
first order.

%P.A.C.S. Numbers: 04.60.-m 04.70.Bw 11.15.-q
\vfil\eject}

\section{1. Introduction}
Since the origin of quantum field theory there have been proposals to add a new scale of length to the theory in order to solve the
problems connected to ultraviolet divergences.
Later, the necessity of introducing a fundamental length scale has also arisen in several attempts to build a theory of quantum gravity.
In these cases, the scale could be identified in a natural way with the Planck length
$L_p=\sqrt{\hbar G\over c^3}\sim1.6\cdot 10^{-35}\;{\rm m}$ [1].

A naive application of the idea of a minimal length, as for example a lattice field theory, would  however break Lorentz invariance.
A way to reconcile the discreteness of spacetime with Lorentz invariance was originally proposed by Snyder [2] a long time ago.
This was the first example of a noncommutative geometry: the length scale should enter the theory through the commutators of
spacetime coordinates, see [3,4]. In particular, the position operators obey the \cor
$$[x_\m,x_\n]=i\b J_\mn,\eqno(1)$$
where $J_\mn$ are the generators of the Lorentz transformations and $\b$ is a parameter of dimension length square that sets the scale
of noncommutativity.\footnote{$^1$}{Throughout this paper we adopt natural units $\hbar=c=1$.}

In more recent times, using ideas coming from the development of noncommutative geometry [5],
the coproduct and star product structures induced by the position operators of the Snyder model have been calculated [6,7].
However, in the Snyder model the algebra of the position operators does not close, as is evident from (1), and hence the bialgebra resulting from the
implementation of the coproduct is not strictly speaking a Hopf algebra, as in other noncommutative geometries. In particular, the
coproduct is not coassociative and the star product is not associative [6].
%For example,
%$$e^{ik\hx}\star e^{iq\hx}=e^{iD(k,q)\hx},\qquad
%D_\m(k,q)={1\over1-\b k\bdot q}\left[\left(1-{\b\,k\bdot q\over1+\rad}\right)k_\m+\rad\, q_\m\right],\eqno(2)$$

A closed Lie algebra can however be obtained if one adds to the position generators the generators of the Lorentz algebra [7].
In this way one can define a proper Hopf algebra, with coassociative coproduct.\footnote{$^2$}{Generally, Lie deformed quantum Minkowski
spaces admit both Hopf algebra and Hopf algebroid structure [8].}  The price to pay is the addition to the formalism of tensorial degrees of
freedom and their conjugate momenta. To distinguish from the standard noncommutative realization of the Snyder model [6],
we call the algebra where the Lorentz generators are added as extended coordinates extended Snyder algebra, and
the theory based on it extended Snyder model.
The physical interpretation of the new degrees of freedom is however not evident, they may be viewed for example as coordinates parametrizing extra
dimensions [7]. %The relation between..

In this paper, we construct new realizations of this extended algebra, perturbatively in the parameter $\b$.
In order to construct them, we define an extended Heisenberg algebra, which includes the Lorentz generators and their conjugate momenta.
Then we consider a Weyl realization of the algebra in terms of the extended Heisenberg algebra, and then generalize it
to the most general one compatible with Lorentz invariance at order $\b$, including the one obtained in [7], and
compute the coproduct and the star product in the general case. We also calculate the twist in the Weyl realization.

We recall here some of the most relevant recent advances in Snyder theory:
in [9] the Snyder algebra was generalized in such a way to maintain the Lorentz invariance;
in [6] the coproduct was calculated, in [7] the same problem was investigated from a geometrical point of view, using the fact that
the momentum space of Snyder can be identified with a coset space;
the twist was investigated in [10,11].
The construction of a field theory was first addressed in [6,7] and then examined in more detail in [12].
Different applications to phenomenology have been considered in [13].
Finally, the extension to a curved background was proposed in [14] and further investigated in [15].
Also the nonrelativistic limit of the theory was studied in a large number of papers, but we shall omit a discussion of this topic.

The paper is organized as follows: in sect.~2 we introduce the extended Snyder model and discuss its Weyl realization in terms
of an extended Heisenberg algebra; in sect.~3 we compute the coproduct and the star product in this realization; in sect.~4 also
the twist is calculated. In sect.~5, generic realizations up to order $\b$ are introduced and coproducts and star products are
obtained. Finally, in sect.~6 the relations of these realizations with that of ref.~[7] and with well-known nonassociative ones are
discussed. In sect.~7 some conclusion are drawn.
\bigbreak
\section{2. Extended Snyder model and Weyl realization}
As mentioned in the introduction, the lack of associativity of the standard realization of the Snyder star product is due to the fact that
this is built in terms of the position coordinates only, whose commutators do not close (cfr.~(1)).
An associative realization of the Snyder model can however be obtained by adding to the algebra generated
by the position coordinates $\hat x_i$ the tensorial coordinates $\hat x_{ij}$, identified with the Lorentz generators, so that
they span the closed algebra (2).
In fact, all Lie-algebra type non commutative spaces induce associative star products and the coproducts of momenta are coassociative.
This implies that the star product we shall obtain in the present framework in eqs.~(26), (27) is associative.
If instead in the star product only the $\cD_i$ were present, without the $\cD_{ij}$, associativity would be lost.

We consider the extended Snyder algebra
%subalgebra of the $N$-dimensional Snyder
algebra generated by the $N$ position operators $\hx_i$ and the
$N(N-1)/2$ antisymmetric Lorentz generators $\hx_\ij$, with $i=0\dots,N-1$,
$$\eqalignno{&[\hx_i,\hx_j]=i\l\b\hx_\ij,\qquad[\hx_\ij,\hx_k]=i\l(\y_{ik}\hx_j-\y_{jk}\hx_i),&\cr
&\ \ [\hx_\ij,\hx_{kl}]=i\l(\y_{ik}\hx_{jl}-\y_{il}\hx_{jk}-\y_{jk}\hx_{il}+\y_{jl}\hx_{ik}),&(2)}$$
where $\l$ and $\b$ are real parameters. In particular, $\b$ can be identified with the Snyder parameter, which is usually assumed to be
of size $L_p^2$, while $\l$ is a dimensionless parameter.
The parameter $\b$ can take both positive and negative values, leading to quite different physical models. However, from an algebraic
point of view both cases can be treated in an essentially unified way. For $\b=0$, the \cor (2) reduce to those of the standard Lorentz algebra
acting on commutative coordinates.

The algebra (2) can be realized in terms of an extended Heisenberg algebra, which includes also the Lorentz generators,
$$\eqalignno{&[x_i,x_j]=[p_i,p_j]=[x_\ij,x_{kl}]=[p_\ij,p_{kl}]=0,&\cr
&[x_i,p_j]=i\y_\ij,\qquad [x_\ij,p_{kl}]=i(\y_{ik}\y_{jl}-\y_{il}\y_{jk}),&\cr
&[x_i,x_{jk}]=[x_i,p_{jk}]=[x_\ij,x_k]=[x_\ij,p_k]=0,&(3)}$$
where $p_i$ and $p_\ij$ are momenta canonically conjugate to $x_i$ and $x_\ij$ respectively, and $p_\ij=-p_{ji}$.
The momenta can be realized in a standard way as
$$p_i=-i{\de\over\de x_i},\qquad p_{ij}=-i{\de\over\de x_{ij}}.\eqno(4)$$
Note that, including the momenta $p_i$ in the algebra (2), with \cor
$$[p_i,p_j]=0,\qquad[\hx_\ij,p_k]=i\l(\y_{ik}p_j-\y_{jk}p_i),\qquad[\hx_i,p_j]=i(\y_\ij+\l^2\b p_ip_j),\eqno(5)$$
one recovers the full original Snyder algebra [2].

To proceed with the computations, it is convenient to exploit the isomorphism between the extended Snyder algebra and $so(1,N)$,
and write the previous formulas more compactly defining, for positive $\b$, $\hx_i\id\sqrt\b\,\hx_{iN}$, $x_i\id\sqrt\b\,x_{iN}$,
$p_i\id p_{iN}/\sqrt\b$, with $\y_{NN}=1$, and $\m=0,\dots N$.\footnote{$^3$}{When $\b<0$ the algebra is isomorphic to $so(2,N-1)$.
The coordinates are defined in the same way, except that the absolute value of $\b$ must be
taken under the square root and $\y_{NN}=-1$. All results are identical, with the appropriate choice of the sign of $\b$.}
The extended Heisenberg algebra (3) becomes then
$$[x_\mn,x_{\r\s}]=[p_\mn,p_{\r\s}]=0,\qquad[x_\mn,p_{\r\s}]=i(\y_{\m\r}\y_{\n\s}-\y_{\m\s}\y_{\n\r}),\eqno(6)$$
while the extended Snyder algebra (2) takes the form
$$[\hx_\mn,\hx_{\r\s}]=i\l C_{\mn,\r\s,\a\b}\,\hx_\ab,\eqno(7)$$
where $C_{\mn,\r\s,\a\b}$ are the structure constants of the $so(1,N)$ algebra,
$$\eqalignno{C_{\mn,\r\s,\a\b}=\ha\Big[&-\y_{\n\r}(\y_{\m\a}\y_{\s\b}-\y_{\s\a}\y_{\m\b})+\y_{\m\s}(\y_{\r\a}\y_{\n\b}-\y_{\n\a}\y_{\r\b})+&\cr
&\ \y_{\m\r}(\y_{\n\a}\y_{\s\b}-\y_{\s\a}\y_{\m\b})-\y_{\n\s}(\y_{\r\a}\y_{\m\b}-\y_{\m\a}\y_{\r\b})\Big].&(8)}$$
that satisfy the symmetry properties $C_{\mn,\r\s,\a\b}=-C_{\n\m,\r\s,\a\b}=-C_{\mn,\s\r,\a\b}=-C_{\mn,\r\s,\b\a}=-C_{\r\s,\mn,\a\b}=-C_{\mn,\a\b,\r\s}$.
%$$[\hx_\mn,\hx_{\r\s}]=i\l(\y_{\m\r}\hx_{\n\s}-\y_{\n\r}\hx_{\m\s}+\y_{\n\s}\hx_{\m\r}-\y_{\m\s}\hx_{\n\r}).\eqno(9)$$

In general, if the \coo $\hat x_\m$ generate a Lie algebra $[\hx_\m,\hx_\n]=iC_{\m\n\l}\hx_\l$
with structure constants $C_{\m\n\l}$, then the universal realization of $\hx_\m$ corresponding to Weyl-symmetric ordering is given by [16]
$$\hx_\m=x_\a\f_{\a\m}(p)=x_\a\({\cal C}\over 1-e^{-{\cal C}}\)_{\m\a},\eqno(9)$$
where ${\cal C}_\mn=C_{\a\m\n}p_\a$.
This realization enjoys the property
$$e^{ik_\m\hx_\m}\triangleright1=e^{ik_\m x_\m},\qquad k_\m\in{\bf R},\eqno(10)$$
where the action $\triangleright$ is given by
$$x_\m\triangleright f(x_\a)=x_\m f(x_\a),\qquad p_\m\triangleright f(x_a)=-i{\de f(x_a)\over\de x_\m},\eqno(11)$$
or, in our case,
$$x_\mn\triangleright f(x_\ab)=x_\mn f(x_\ab),\qquad p_\mn\triangleright f(x_\ab)=-i{\de f(x_\ab)\over\de x_\mn}=[p_\mn, f(x_\ab)],\eqno(12)$$
Hence, the corresponding Weyl realization of $\hx_\mn$ in terms of the extended Heisenberg algebra (6) reads [16]
$$\hx_\mn=x_{\a\b}\({\l\,{\cal C}\over 1-e^{-\l\cal C}}\)_{\mn,\ab}=x_\mn+{\l\over2}x_\ab\,{\cal C}_{\mn,\ab}+{\l^2\over12}x_\ab\({\cal C}^2\)_{\mn,\ab}
+{\cal O}(\l^4).\eqno(13)$$
where
$$\eqalignno{&{\cal C}_{\mn,\ab}=\ha\,C_{\r\s,\mn,\ab}p_{\r\s}=\ha(\y_{\m\a}p_{\n\b}-\y_{\m\b}p_{\n\a}+\y_{\n\b}p_{\m\a}-\y_{\n\a}p_{\m\b}),&\cr
&\({\cal C}^2\)_{\mn,\ab}=\ha\sum_{k=0}^2\pmatrix{2\cr k\cr}\Big((p^k)_{\m\a}(p^{2-k})_{\n\b}-(p^{2-k})_{\m\b}(p^k)_{\n\a}\Big),&(14)}$$
and $p_\mn$ is written in matricial notation.

Inserting $\cal C$ in (13), we find up to order $\l^2$,
$$\hat x_\mn=x_\mn+{\l\over 2}(x_{\m\a}p_{\n\a}-x_{\n\a}p_{\m\a})-{\l^2\over 12}(x_{\m\a}p_{\n\b}p_{\a\b}-x_{\n\a}p_{\m\b}p_{\a\b}-2x_{\a\b}p_{\m\a}p_{\n\b}).
\eqno(15)$$
One has then
$$\eqalignno{[\hx_\mn,p_{\r\s}]&=i(\y_{\m\r}\y_{\n\s}-\y_{\m\s}\y_{\n\r})+{i\l\over2}(\y_{\m\r}p_{\n\s}-\y_{\n\r}p_{\m\s}+\y_{\n\s}p_{\m\r}-\y_{\m\s}p_{\n\r})&\cr
&-{i\l^2\over 12}(\y_{\m\r}p_{\n\a}p_{\s\a}-\y_{\m\s}p_{\n\a}p_{\r\a}-\y_{\n\r}p_{\m\a}p_{\s\a}+\y_{\n\s}p_{\m\a}p_{\r\a}+2p_{\m\r}p_{\n\s}-2p_{\n\r}p_{\m\s}).&(16)}$$

One can rewrite eq.~(15) in terms of its components as
$$\eqalignno{\hx_i&=x_i+{\l\over2}\big(x_kp_{ik}-\b x_{ik}p_k\big)-{\l^2\over12}\big(x_\k p_{kl}p_{il}+\b(-x_k p_kp_i+x_ip_k^2-x_{ik}p_lp_{kl}-2x_{kl}p_kp_{il})\big),&\cr
\hx_\ij&=x_\ij+{\l\over2}\big(x_ip_j+x_{ik}p_{jk}-(i\leftrightarrow j)\big)-{\l^2\over12}\big(x_{ik}p_{jl}p_{kl}-x_{kl}p_{ik}p_{jl}-x_ip_kp_{jk}+2x_kp_ip_{jk}&\cr
&+\b x_{ik}p_kp_j-(i\leftrightarrow j)\big).&(17)}$$

In the limit $\l\b=L_p^2$, $\l=0$, the algebra (2) becomes the DFR (Moyal) algebra [3] and the realization (15) takes the form
$$\hx_i=x_i-{L_p^2\over2}\,x_{ik}p_k,\qquad\hx_\ij=x_\ij.\eqno(18)$$
The corresponding Lorentz generators are
$$M_\ij=x_ip_j-x_jp_i+x_{ik}p_{jk}-x_{jk}p_{ik}.\eqno(19)$$
\vfil\eject
\section{3. Coproduct and star product in Weyl realization}
In order to compute the coproduct of the Hopf algebra, we use the formalism introduced in [17].
We define a function $\cP_\mn(tk_\ab)$ that satisfies the differential equation
$${d\cP_\mn\over dt}={i\over2}[p_\mn,k_{\r\s}\hx_{\r\s}]\big|_{p\to\cP(tk)}=k_{\r\s}\F_{\mn,\r\s}(\cP_{\a\b}),\eqno(20)$$
with initial condition $\cP_\mn(0)=q_\mn$. The function $\F_{\mn,\r\s}(p_{\a\b})$ is defined from (15) as
$\hat x_\mn=x_{\r\s}\F_{\r\s,\mn}$.
In our case, equation (20) takes the form
$${d\cP_\mn\over dt}=k_\mn-{\l\over2}(k_{\m\a}\cP_{\n\a}-k_{\n\a}\cP_{\m\a})-
{\l^2\over12}(k_{\m\a}\cP_{\a\b}\cP_{\n\b}-k_{\n\a}\cP_{\a\b}\cP_{\m\b}-2k_{\a\b}\cP_{\m\a}\cP_{\n\b}),\eqno(21)$$
and with the given initial condition has solution
$$\eqalignno{\cP_\mn&=q_\mn+tk_\mn-{\l t\over2}\Big(k_{\m\a}q_{\n\a}-k_{\n\a}q_{\m\a}\Big)
-{\l^2\over12}\Big(\big(k_{\m\a}q_{\a\b}q_{\n\b}-k_{\n\a}q_{\a\b}q_{\m\b}-2k_{\a\b}q_{\m\a}q_{\n\b}\big)t&\cr
&+\big(k_{\m\a}k_{\a\b}q_{\n\b}-k_{\n\a}k_{\a\b}q_{\m\b}-2k_{\m\a}k_{\n\b}q_{\a\b}\big)t^2\Big).&(22)}$$
We can now define $P_\mn(k_\mn,q_\mn)\id\cP_\mn(t=1)$, so that
$$\eqalignno{P_\mn(k_\mn,q_\mn)&=\ k_\mn+q_\mn-{\l\over2}\Big(k_{\m\a}q_{\n\a}-k_{\n\a}q_{\m\a}\Big)
-{\l^2\over12}\Big(k_{\m\a}q_{\a\b}q_{\n\b}-k_{\n\a}q_{\a\b}q_{\m\b}-2k_{\a\b}q_{\m\a}q_{\n\b}&\cr
&+k_{\m\a}k_{\a\b}q_{\n\b}-k_{\n\a}k_{\a\b}q_{\m\b}-2k_{\m\a}k_{\n\b}q_{\a\b}\Big).&(23)}$$
Defining then $\cK_\mn(k_\mn)\id P_\mn(q_\mn=0)$, one has $\cK_\mn=k_\mn$, and therefore also its inverse function $\cK^\mo_\mn(k_\mn)=k_\mn$.

It can be shown that the generalized momentum addition law is given by [17]
$$k_\mn\oplus q_\mn\id\cD_\mn(k_\ab,q_\ab)=P_\mn(\cK^\mo_\ab,q_\ab),\eqno(24)$$
and hence in our case $\cD_\mn(k_\ab,q_\ab)=P_\mn(k_\ab,q_\ab)$.
This yields the coproduct
$$\eqalignno{\D p_\mn&=\cD_\mn(p_\mn\otimes1,1\otimes p_\mn)=\ \D_0p_\mn-{\l\over2}\Big(p_{\m\a}\otimes p_{\n\a}-p_{\n\a}\otimes p_{\m\a}\Big)-{\l^2\over12}\Big(p_{\m\a}\ot p_{\a\b}p_{\n\b}&\cr&
-p_{\n\a}\ot p_{\a\b}p_{\m\b}-2p_{\a\b}\ot p_{\m\a}p_{\n\b}+p_{\m\a}p_{\a\b}\ot p_{\n\b}-p_{\n\a}p_{\a\b}\ot p_{\m\b}-2p_{\m\a}p_{\n\b}\ot p_{\a\b}\Big),&(25)}$$
with $\D_0p_\mn=p_\mn\otimes1+1\otimes p_\mn$. It is straightforward to explicitly check the coassociativity of this coproduct.
It is also easy to see that the antipode is trivial, $S(p_\mn)=-p_\mn$.

Recalling our definitions $\hx_i=\sqrt\b\hx_{iN}$ and $p_i=p_{iN}/\sqrt\b$, we can write the functions $\cD_\ab$ in terms of their components, namely
$$\eqalignno{\cD_i(k,q)=&\ k_i+q_i-{\l\over2}\Big[k_jq_\ij-k_{ij}q_j\Big]+{\l^2\over12}\Big[\b(k_ik_jq_j-k_j^2q_i)-k_jk_{jk}q_{ik}+2k_{ik}k_jq_{jk}&\cr
&+k_\ij k_{jk}q_k+\b(k_jq_jq_i-k_iq_j^2)+k_{ij}q_{jk}q_k-2k_{jk}q_kq_{ij}-k_jq_{jk}q_{ik}\Big],&(26)}$$

$$\eqalignno{\cD_\ij(k,q)=&\ k_\ij+q_\ij-{\l\over2}\Big[k_{ik}q_{jk}+\b k_iq_j-(i\lra j)\Big]+{\l^2\over12}\Big[k_{ik}k_{jl}q_{kl}-k_{ik}k_{kl}q_{jl}&\cr
&+\b(k_ik_kq_{jk}-k_{ik}k_kq_j-2k_ik_{jk}q_k)+k_{kl}q_{ik}q_{jl}-k_{ik}q_{kl}q_{jl}&\cr
&-\b(k_{ik}q_kq_j-k_iq_kq_{jk}-2k_kq_iq_{jk})-(i\lra j)\Big].&(27)}$$
The functions $\cD(q,k)$ satisfy the symmetry properties
$$\cD_i(q,k)\big|_\l=\cD_i(k,q)\big|_{-\l},\qquad\cD_\ij(q,k)\big|_\l=\cD_\ij(k,q)\big|_{-\l}.\eqno(28)$$

It also holds
$$e^{{i\over2}k_{\mn}\hx_{\mn}} e^{{i\over2}q_\rs\hx_{\rs}}=e^{{i\over2}\cD_\mn(k,q)\hx_\mn},\eqno(29)$$
and
$$e^{{i\over2}k_\mn x_\mn}\star e^{{i\over2}q_\rs x_\rs}=e^{{i\over2}k_{\mn}\hx_{\mn}} e^{{i\over2}q_\rs\hx_{\rs}}\tr1=e^{{i\over2}\cD_\mn(k,q)\hx_\mn}\tr1=e^{{i\over2}\cD_\mn(k,q)x_\mn}.\eqno(30)$$
Moreover, we can write
$$\eqalignno{&e^{{i\over2}k_{\mn}\hx_{\mn}}=e^{ik_i\hx_i+{i\over2}k_{\ij}\hx_{\ij}},&\cr
&e^{ik_ix_i+{i\over2}k_{\ij}x_{\ij}}\star e^{iq_kx_k+{i\over2}q_{kl}x_{kl}}=e^{i\cD_ix_i+{i\over2}\cD_\ij x_{\ij}}.&(31)}$$

In particular, from (26) and (27) one can obtain the star product for plane waves. Notice that the star product of two translations clearly will have a component also in the direction of
rotations,
$$\eqalignno{e^{ik_ix_i}\star e^{iq_jx_j}=&\ e^{i\[k_i+q_i-{1\over12}\l^2\b(q_j^2k_i-k_jq_jq_i+k_j^2q_i-k_jq_jk_i)\]x_i-{i\over2}\l\b k_iq_jx_{ij}},&\cr
e^{{i\over2}k_{ij}x_{ij}}\star e^{{i\over2}q_{kl}x_{kl}}=&\ e^{{i\over2}\[k_{ij}+q_{ij}-\l k_{ik}q_{jk}-{1\over6}\l^2(k_{ik}q_{kl}q_{jl}-k_{kl}q_{ik}q_{jl}+k_{ik}k_{kl}q_{jl}-k_{ik}k_{jl}q_{kl})\]x_{ij}},&\cr
e^{ik_kx_k}\star e^{{i\over2}q_{ij}x_{ij}}=&\ e^{i\[k_i-{\l\over2}k_jq_{ij}-{1\over12}\l^2k_jq_{jk}q_{ik}\]x_i+{i\over2}\[q_\ij+{1\over6}\l^2\b k_ik_kq_{jk}\]x_{ij}},&\cr
e^{{i\over2}k_{ij}x_{ij}}\star e^{{i\over2}q_kx_k}=&\ e^{i\[q_i+{\l\over2}k_{ij}q_j+{1\over12}\l^2k_{ij}k_{jk}q_k\]x_i+{i\over2}\[k_\ij-{1\over6}\l^2\b k_{ik}q_kq_j\]x_{ij}}.&(32)}$$
This star product is associative.
One can also check that the star products of the coordinates $x_i$ and $x_{ij}$ satisfy the extended Snyder algebra. In fact, according to [7],
denoting $k$ the vector $k_i$, $\bl$ the tensor $l_{ij}$ and so on, and defining $e_{k,\bl}=e^{k_ix_i+\bl_{jk}x_{jk}}$, the star
product of the coordinates can be evaluated as follows:
$$\eqalignno{x_i\star x_j=&\int dk\,dq\,d\bl\,d\br\,\d(k)\d(q)\d(\bl)\d(\br)\de_{k_i}\de_{q_j}(e_{k,\bl}\star e_{q,\br})=\hx_i\triangleright x_j=x_ix_j+i{\l\b\over2}x_{ij},&\cr
x_{ij}\star x_{kl}=&\int dk\,dq\,d\bl\,d\br\,\d(k)\d(q)\d(\bl)\d(\br)\de_{\bl_{ij}}\de_{\br_{kl}}(e_{k,\bl}\star e_{q,\br})=\hx_{ij}\triangleright x_{kl}=x_{ij}x_{kl}
+i{\l\over2}\Big(\y_{ik}x_{jl}-\y_{jk}x_{il}&\cr&-\y_{il}x_{jk}+\y_{jl}x_{ik}\Big),&\cr
x_k\star x_{ij}=&\int dk\,dq\,d\bl\,d\br\,\d(k)\d(q)\d(\bl)\d(\br)\de_{k_k}\de_{\br_{ij}}(e_{k,\bl}\star e_{q,\br})=\hx_k\triangleright x_{ij}=x_kx_{ij}-i{\l\over2}\Big(\y_{ik}x_j-\y_{jk}x_i\Big),&\cr
x_{ij}\star x_k=&\int dk\,dq\,d\bl\,d\br\,\d(k)\d(q)\d(\bl)\d(\br)\de_{\bl_{ij}}\de_{q_k}(e_{k,\bl}\star e_{q,\br})=\hx_{ij}\triangleright x_k=x_{ij}x_k+i{\l\over2}\Big(\y_{ik}x_j-\y_{jk}x_i\Big).&(33)}$$
Therefore,
$$\eqalignno{&[x_i,x_j]_\star=i\l\b x_{ij},\qquad[x_{ij},x_k]_\star=i\l(\y_{ik}x_j-\y_{jk}x_i),&\cr
&\ \ [x_{ij},x_{kl}]_\star=i\l(\y_{ik}x_{jl}-\y_{jk}x_{il}-\y_{il}x_{jk}+\y_{jl}x_{ik}),&(34)}$$
which is isomorphic to the algebra (2).

\bigbreak
\section{4. The twist for the Weyl realization}
In this section, we construct the twist operator at second order in $\l$, using a perturbative approach.
The twist is defined as a bilinear operator such that $\D h=\cF\D_0h\cF^\mo$ for each $h\in so(1,N)$.

The twist in a Hopf algebroid sense can be computed by means of the formula [10,18]
$$\cF^\mo\id e^F=e^{-{i\over2}p_\mn\ot x_\mn}e^{{i\over2}p_\rs\ot\hx_\rs}.\eqno(35)$$
By the Baker-Campbell-Haussdorf formula  $e^Ae^B=e^{A+B+\ha[A,B]+\dots}$, one gets
$$F={i\over2}\,p_{\m\n}\ot(\hx_\mn-x_\mn)-{1\over8}\,p_\mn p_\rs\ot[x_\mn,\hx_\rs]+\dots\eqno(36)$$
where we can safely ignore further terms because it can be explicitly checked that they give contributions of order $\l^3$.

Substituting (15) in (36), one obtains
$$F={i\l\over2}p_{\a\g}\ot x_{\a\b}p_{\g\b}-{i\l^2\over24}\big(2p_{\a\g}\ot x_{\a\b}p_{\b\d}p_{\g\d}-2p_{\g\d}\ot x_{\a\b}p_{\a\g}p_{\b\d}
-p_{\a\g}p_{\b\d}\ot x_{\a\b}p_{\g\d}+p_{\a\g}p_{\d\g}\ot x_{\a\b}p_{\d\b}\big).\eqno(37)$$

Using the Hadamard formula $e^ABe^{-A}=B+[A,B]+\ha[A,[A,B]]+\dots$, it is easy to check that
$$\cF\D_0p_\mn\cF^\mo=\D p_\mn,\eqno(38) $$
with $\D p_\mn$ given in (25), as expected.

\section{5. Generic realizations}
We consider now the most general realization of the \cor (2) in terms of the elements of the extended Heisenberg algebra (3), up to second order in $\l$.
Of course, this will deform the \cor between coordinates and momenta in (5).

The generic form of the Lorentz-covariant combinations of the generators of the algebra (3), linear in $x_i$, $x_\ij$, up to order $\l^2$ is given
by\footnote{$^4$}{In principle, one may add further terms to (39), namely the terms  $x_i p_{kl}p_{kl}$ and $x_{kl} p_{kl} p_i$ to $\hat x_i$, and
$x_{ij} p_k p_k$,  $x_{ij} p_{kl} p_{kl}$,  $x_{kl} p_{kl} p_{ij}$, $x_k p_k p_{ij}$ to $\hat x_{ij}$. However these terms must vanish if one requires
that the Snyder algebra is satisfied.}
$$\eqalignno{\hat x_i&=x_i+\l\big(\b c_0x_{ik}p_k+c_1x_kp_{ik}\big)+\l^2\big(\b(c_2x_ip_k^2+c_3x_kp_kp_i+c_4x_{ik}p_{kl}p_l+c_5x_{kl}p_kp_{il})+c_6x_kp_{kl}p_{il}\big),&\cr&\cr
\hat x_\ij&=x_\ij+\l\big(d_0x_{ik}p_{jk}+d_1x_ip_j-(i\lra j)\big)+\l^2\big(\b d_2x_{ik}p_kp_j+d_3x_{ik}p_{kl}p_{jl}+d_4x_{kl}p_{ik}p_{jl}+d_5x_ip_kp_{jk}&\cr
&\quad+d_6x_kp_{ik}p_j-(i\lra j)\big).&(39)}$$
In order to satisfy (2) to first order in $\l$ one must have
$$c_0=-\ha,\qquad d_0=\ha,\qquad c_1+d_1=1.\eqno(40)$$
Hence, at this order one has one free parameter. In particular, in the Weyl realization (17), $d_1=c_1=\ha$.

To second order in $\l$, one has ten new parameters $c_2,\dots,c_6$, $d_2,\dots,d_6$ that must satisfy the six independent relations
$$\eqalignno{&\ {c_1\over2}-2c_2+c_3=d_1,\qquad{c_1\over2}+c_4+c_5=\ha,\qquad d_3-2d_4=-{1\over4},&\cr
&c_5-d_2={1\over4},\qquad{c_1\over2}+c_6-d_6=0,\qquad{c_1\over2}-c_1d_1+c_6+d_5=0.&(41)}$$
Hence  up to second order one has five free parameters. For example, one may choose as free parameters $c_1$, $c_2$, $c_4$, $d_4$ and $d_5$, so that $d_1=1-c_1$ and
$$\eqalignno{&c_3=1-{3c_1\over2}+2c_2,\quad c_5=\ha-{c_1\over2}-c_4\quad c_6={c_1\over2}-c_1^2-d_5,&\cr
&\ \ d_2=\qu-{c_1\over2}-c_4,\quad d_3=-{1\over4}+2d_4,\quad d_6=c_1-c_1^2-d_5.&(42)}$$
It is easy to verify that the coefficients of the Weyl realization (17) satisfy the  above relations with $c_1=\ha$, $c_2=-c_4=-d_4=-d_5=-{1\over12}$.

Note that setting $\b=0$ in (39) one obtains realizations of the \poi algebra. For example, the Weyl realization for the operators $\hat x_i$ and $\hat x_\ij$
of the \poi algebra becomes
$$\eqalignno{\hat x_i&=x_i+{\l\over2}\,x_kp_{ik}-{\l^2\over12}\,x_kp_{kl}p_{il},&\cr
\hx_\ij&=x_\ij+\[{\l\over2}\big(x_ip_j+x_{ik}p_{jk}\big)-{\l^2\over12}\big(x_{ik}p_{jl}p_{kl}-x_{kl}p_{ik}p_{jl}-x_ip_kp_{jk}+2x_kp_ip_{jk}\big)
-(i\leftrightarrow j)\].&(43)}$$

Through the same procedure as in the previous section, one can determine the coproduct for the generic realization (39). The differential equations for
$\cP_i(tk)$ and $\cP_\ij(tk)$ are
$$\eqalignno{{d\cP_i\over dt}&=i\Big[p_i,k_k\hx_k+\ha k_{kl}\hx_{kl}\Big]\bigg|_{p\to\cP(tk)},&\cr
{d\cP_\ij\over dt}&=i\Big[p_\ij,k_k\hx_k+\ha k_{kl}\hx_{kl}\Big]\bigg|_{p\to\cP(tk)},&(44)}$$
with initial conditions $\cP_i(0)=q_i$ and $\cP_\ij(0)=q_\ij$. After some calculations, one can write down the functions $\cD_i(k,q)$ and $\cD_\ij(k,q)$
that appear in the star product of plane waves,
$$\eqalignno{\cD_i(k,q)&=k_i+q_i+\l\(-c_1k_jq_\ij+d_1k_{ij}q_j\)+{\l^2\over2}\Big[\b\(c_0c_1+c_3\)k_j^2q_i+\b(-c_0c_1+2c_2+c_3)k_ik_jq_j&\cr
&+\(c_1^2-c_1d_0-c_1d_1+c_6+d_6\)k_jk_{jk}q_{ik}+(c_1d_0+c_1d_1+c_6-d_5)k_{ik}k_jq_{jk}&\cr
&+\(d_1^2+d_5-d_6\)k_\ij k_{jk}q_k+2\b c_2k_iq_j^2+2\b c_3k_jq_jq_i+2d_5k_{ij}q_{jk}q_k+2d_6k_{jk}q_jq_{ik}&\cr
&+2c_6k_jq_{jk}q_{ik}\Big],&(45)}$$
and
$$\eqalignno{\cD_\ij(k,q)&=k_\ij+q_\ij+\l\(-d_0k_{ik}q_{jk}+\b c_0k_iq_j-(i\lra j)\)+{\l^2\over2}\Big[\b\(-c_0c_1+c_4-c_5\)k_ik_kq_{jk}&\cr
&+(-d_0^2+d_3+2d_4)k_{ik}k_{kl}q_{jl}+\(d_0^2+d_3\)k_{ik}k_{jl}q_{kl}+\b\(c_0d_0+c_5+d_2\)k_{ik}k_kq_j&\cr
&+\b(c_0d_0+c_0d_1+c_4-d_2)k_ik_{jk}q_k+2\b d_2k_{ik}q_kq_j+2d_3k_{ik}q_{kl}q_{jl}+2d_4k_{kl}q_{ik}q_{jl}&\cr
&+2\b c_4k_iq_kq_{jk}+2\b c_5k_kq_{ik}q_j-(i\lra j)\Big].&(46)}$$
From these functions one can easily obtain the star product and the coproduct in the general case, see (25) and (31). In particular, for $c_0=-\ha$ and $k_\ij=q_\ij=0$, one has
$$e^{ik_ix_i}\star e^{iq_jx_j}=e^{i\[k_i+q_i+{1\over2}\l^2\b\(2c_2q_j^2k_i+2c_3k_jq_jq_i+(c_3-{c_1\over2})k_j^2q_i+(2c_2+c_3+{c_1\over2})k_jq_jk_i\)\]x_i-{i\over2}\l\b k_iq_jx_{ij}},
\eqno(47)$$
which for $c_1=\ha$, $c_2=-c_3=-{1\over12}$ reduces to the first relation in (32).
%\vfil\eject
\section{6. Comparison with the Girelli-Livine approach}
The authors of [7] studied our model in 3D Euclidean space using geometric methods, with a very different parametrization, adapted to
the coset space nature of the Snyder momentum space.
In our notations, their star product for plane waves takes the form, at second order in $\l$,
$$\eqalign{e^{ik_ix_i}\star e^{iq_jx_j}
%&=\exp\[i\(k_i+q_i+{\l\b\over2}(k_jq_jk_i+k_j^2q_i+2k_jq_jq_i)\)x_i\]\exp\[i{\l\b\over8}(k_iq_j-k_jq_i)x_{ij}\]\cr&
=\exp\[i\(k_i+q_i+{\l^2\b\over2}(k_jq_jk_i+k_j^2q_i+2k_jq_jq_i)\)x_i-i{\l\b\over2}k_iq_jx_{ij}\].}\eqno(48)$$
This expression corresponds to a realization (39) with $c_0=-\ha$, $d_0=\ha$ and $c_1=c_2=0$. It follows from (42) that $c_3=1$, but the other coefficients are not determined and
depend on three free parameters.
If one also requires $d_5=0$, this may be called a generalized Snyder realization, since it obeys all the commutation relations of the original Snyder model [2],
given by (2) and (5).
Note that the momenta $p_\ij$ do not appear in these relations. Of course, additional \cor are obeyed by the momenta $p_\ij$, but they are not of interest for our
considerations.

One may consider more general realizations belonging to the previous class, with $c_0=-\ha$, $d_0=\ha$, $c_1=0$ and three free parameters. For example, $c_2=-\ha$, implies $c_3=0$
and gives rise to a realization that, for $d_5=0$, reproduces at order $\b$ the \cor of the Maggiore realization introduced in [9].

More generally, these representations generalize those introduced in [10], with arbitrary $c_2$ and $c_3=1+2c_2$. In particular, one can choose the free parameters
such that
$$\eqalignno{\hx_i&=x_i+{\l^2\b\over2}\Big[(c_3-1)x_ip_k^2+2c_3x_kp_kp_i\Big]-{\l\b\over2}\hm_{ik}p_k&\cr
\hx_\ij&=\hm_\ij+\l(x_ip_j-x_jp_i),&(49)}$$
where the $\hm_\ij$ generate the Lorentz algebra $so(1,N-1)$ and
$$[\hm_\ij,x_k]=[\hm_\ij,p_k]=0.\eqno(50)$$
For example, in the Weyl realization of $\hm_\ij$, $d_3=-d_4=-{1\over12}$, leaving $c_3$ as a free parameter.
In the limit $\b=0$, $\hx_i$ reduces to $x_i$.
%$$\hm_\ij=x_\ij+{\l\over2}(x_{ik}p_{jk}-x_{ik}p_{jk})+{\l^2\over12}(-x_{ik}p_{kl}p_{jl}+x_{jk}p_{kl}p_{il}+2x_{kl}p_{ik}p_{jl}).\eqno(45)$$
%and they can be interpreted as the Weyl realization of the Lorentz algebra. This choice corresponds to $c_4=c_5={1\over4}$, $d_4=-d_3={1\over12}$ and $d_5=d_6=c_6=0$.

\section{7. Conclusions}
The coalgebra usually associated to the Snyder model is noncoassociative, and this fact prevents the definition of a proper Hopf algebra,
whose coproduct is by definition coassociative.
The reason is that the algebra of the position operators of the Snyder model does not close. However this can be remedied by including the Lorentz generators
in the defining algebra [7]. In this way a standard coassociative Hopf algebra can be defined.

In this paper we have studied the realizations of this extended algebra in terms of the deformations of an extended Heisenberg algebra, which contains tensorial
elements that in the deformation assume the role of Lorentz generators.
We have obtained the coproduct, the star product and the twist in the case of a Weyl realization. We have also considered the most general realization
of the algebra up to second order in the expansion parameter $\l$ (or equivalently at first order in the Snyder parameter $\b$) and calculated the corresponding coproduct
and star product.

Although this approach may be considered more rigorous than the standard one from a mathematical point of view, the physical interpretation of the new degrees of freedom,
related to the Lorentz generators and their momenta, is still an issue.
In ref.~[7] the tensorial coordinates $x_\ij$ were interpreted in a Kaluza-Klein perspective, as coordinates of extra dimensions, hence not identified with Lorentz generators.
It is also important to note that the action of noncommutative tensorial coordinates on 1 is defined to give commutative tensorial coordinates (see eqs.~(30) and (33)).
The noncommutative tensorial coordinates are related to the parametrization of the dual Lorentz group. This topic is presently being investigated.

In applications, one may for example build a field theory assuming that the fields $\q(\hx_\mn)$ depend only on the spacetime coordinates [7], \ie $\q(x_\mn)=\f(\hx_i)\d(\hx_\ij)$.
In this way one would however recover the usual nonassociative star product. Another possibility is that the extra coordinates parametrize a compactified
internal space. In this case associativity would be preserved, but nontrivial physical consequences would presumably arise. We leave the investigation of this possibility
for future work.
In any case, a field theory based on this formalism could avoid the shortcomings due to the nonassociativity of the star product [12], but
different problems can arise because of the intertwining between the position and the extra degrees of freedom [7].

To conclude, we observe that also the standard commutative theory, as well as DFR spacetime [3], can be formulated in this extended framework, as we have observed several times in the text.
The investigation of these elementary cases could be a good starting point to better understand the physical implications of the present formalism, in particular in relation with
quantum field theory.
\vfill\eject

\beginref
\ref[1] L.J. Garay, \IJMP{A10}, 145 (1995).

\ref[2] H.S. Snyder, \PR{71}, 38 (1947).

\ref[3] S. Doplicher, K. Fredenhagen and J. E. Roberts, \PL{B331}, 39 (1994); S. Doplicher, K. Fredenhagen and J. E. Roberts, \CMP{172}, 187 (1995).

\ref[4] G. Amelino-Camelia, J. Lukierski and A. Nowicki, Phys. At. Nucl. {\bf 61}, 1811 (1998); G. Amelino-Camelia, and M. Arzano, \PR{D65}, 084044 (2002).

\ref[5] S. Majid, {\it Foundations of quantum group theory}, Cambridge Un. Press 1995.

\ref[6] M.V. Battisti and S. Meljanac, \PR{D82}, 024028 (2010).

\ref[7] F. Girelli and E. Livine, \JHEP{1103}, 132 (2011).

\ref[8] J. Lukierski, D. Meljanac, S. Meljanac, D. Pikuti\'c and M. Woronowicz, \PL{B777}, 1 (2018); J. Lukierski, S. Meljanac and M. Woronowicz, \PL{B789}, 82 (2019).

\ref[9] M.V. Battisti and S. Meljanac, \PR{D79}, 067505 (2009).

\ref[10] S. Meljanac, D. Meljanac, S. Mignemi and R. \v Strajn, \IJMP{A32}, 1750172 (2017).

\ref[11] S. Meljanac, D. Meljanac, S. Mignemi, D. Pikuti\'c and R. \v Strajn, \EPJ{C78}, 194 (2018).

\ref[12] S. Meljanac, D. Meljanac, S. Mignemi and R. \v Strajn, \PL{B768}, 321 (2017);
S. Meljanac, S. Mignemi, J. Trampeti\'c and J. You, \PR{D96}, 045021 (2017);
S. Meljanac, S. Mignemi, J. Trampeti\'c and J. You, \PR{D97}, 055041 (2018);
A. Franchino-Vi\~nas and S. Mignemi, \PR{D98}, 065010 (2018).

\ref[13] S. Mignemi and R. \v Strajn, \PR{D90}, 044019 (2014);
S. Mignemi and A. Samsarov, \PL{A381}, 1655 (2017);
S. Mignemi and G. Rosati, \CQG{35}, 145006 (2018).

\ref[14] J. Kowalski-Glikman and L. Smolin, \PR{D70}, 065020 (2004).

\ref[15] H.G. Guo, C.G. Huang and H.T. Wu, \PL{B663}, 270 (2008);
R. Banerjee, K. Kumar and D. Roychowdhury, \JHEP{1103}, 060 (2011);
S. Mignemi, \CQG{26}, 245020 (2009).

\ref[16] N. Durov, S. Meljanac, A. Samsarov and Z. \v Skoda, J. Algebra {\bf 309}, 318 (2007);
S. Meljanac, T. Martini\' c-Bila\' c and S.~Kre\v si\' c-Juri\' c, \JMP{61}, 051705 (2020).

\ref[17] S. Meljanac, D. Meljanac, A. Samsarov and M. Stoji\'{c}, \MPL{A25}, 579 (2010);
S. Meljanac, D. Meljanac, A. Samsarov and M. Stoji\'{c}, \PR{D83}, 065009 (2011).

\ref[18] S. Meljanac, D. Meljanac, F. Mercati, D. Pikutic, \PL{B766}, 181 (2017);
S. Meljanac, D. Meljanac, A. Pachol, D. Pikuti\'c, \JoP{A50}, 265201 (2017);
D. Meljanac, S. Meljanac, S. Mignemi, R. \v Strajn, \PR{D99}, 126012 (2019).
%D. Meljanac, S. Meljanac, Z. \v Skoda, R. \v Strajn, \arx{1911.03967}.

\endref
\end